\documentclass[11pt,a4paper]{article}
\usepackage{jheppub_kim}

\usepackage{pdflscape}
\usepackage{amsmath}
\usepackage{amssymb}
\usepackage{dcolumn}
\usepackage{bm}
\usepackage{color}
\usepackage{epsfig}
\usepackage{amsfonts}
\usepackage{graphicx}
\usepackage{subfigure}
\usepackage{dcolumn}

\newcommand{\be}{\begin{equation}}
\newcommand{\ee}{\end{equation}}
\newcommand{\bea}{\begin{eqnarray}}
\newcommand{\eea}{\end{eqnarray}}

\def\be{\begin{equation}}
\def\ee{\end{equation}}
\def\bea{\begin{eqnarray}}
\def\eea{\end{eqnarray}}

\begin{document}

\title{Unified universe history through phantom extended Chaplygin gas}

\author{B. Pourhassan}

\affiliation{School of Physics, Damghan University, Damghan, Iran}

\emailAdd{b.pourhassan@du.ac.ir}

\abstract{We investigate the universe evolution from inflation to late-time acceleration in a
unified way, using a two-component fluid  constituted from extended Chaplygin gas
alongside a phantom scalar field. We extract solutions for the various cosmological
eras, focusing on the behavior of the scale factor, the various density parameters and
the equation-of-state parameter. Furthermore, we extract and discuss bouncing solutions.
Finally, we examine the perturbations of the model, ensuring about their stability and
extracting the predictions for the tensor-to-scalar ratio.}


\maketitle

\section{Introduction}

Cosmological observations verify the accelerating expansion of the universe,
indicating that the transition from deceleration to acceleration was realized
in the recent cosmological past \cite{Riess:1998cb,Perlmutter:1998np}. The
reasonable explanation would be the presence of a simple cosmological
constant, however the possible dynamical features have led to two main
alternative directions. The first is to modify the gravitational sector  acquiring a
modified cosmological dynamics, as in $f(R)$
gravity \cite{Starobinsky:1980te,Nojiri:2006gh,Bertolami:2007gv,Cai:2010zma}, in
Gauss-Bonnet gravity
\cite{Wheeler:1985nh,Nojiri:2005jg}, in Lovelock gravity
\cite{Lovelock:1971yv,Deruelle:1989fj}, in Weyl gravity
\cite{Mannheim:1988dj,Flanagan:2006ra}, in Ho\v{r}ava-Lifshitz gravity
\cite{Horava:2008ih,Horava:2009uw,Kiritsis:2009sh,Saridakis:2009bv,Saridakis:2012ui}, in nonlinear massive gravity
\cite{deRham:2010kj,Hinterbichler:2011tt,deRham:2014zqa,Leon:2013qh}, in $f(T)$ gravity
\cite{Ferraro:2008ey,Linder:2010py,Chen:2010va,Dent:2011zz,Bamba:2010wb,
Capozziello:2012zj,
Li:2013xea,Ong:2013qja,Kofinas:2014owa,Kofinas:2014daa}, etc (for a review see \cite{Capozziello:2011et}). The second way is to modify the universe
content, introducing the dark energy sector, with its simpler candidates being a canonical
(quintessence) scalar field
\cite{Ratra:1987rm,Wetterich:1987fm,Liddle:1998xm,Guo:2006ab,Dutta:2009yb}, a
phantom field
\cite{Caldwell:1999ew,Caldwell:2003vq,Nojiri:2003vn,Onemli:2004mb,
Saridakis:2008fy,Saridakis:2009pj,Gupta:2009kk}, the combination of both
fields in a unified (quintom)
scenario \cite{Guo:2004fq,Zhao:2006mp,Cai:2009zp}, or more complicated
models such as in K-essence \cite{ArmendarizPicon:2000ah}, Hordenski
\cite{Horndeski} and generalized Galileon
theories \cite{DeFelice:2010nf,Deffayet:2011gz,Leon:2012mt} (for a review on dark energy
see \cite{Copeland:2006wr}). Finally, since the dynamical nature of dark
energy introduces the ``coincidence problem'', namely why are the current
dark energy and matter densities of the same order although they evolve
differently, many extensions of the above scenarios were developed, with an
explicit interaction between the dark energy and dark matter sectors
\cite{P45,P54,P49,P58,P48,P74,Boehmer:2008av,P61,Chen:2008ft,P50}.

Since in the above scenarios the dark matter and dark energy belong to two
different sectors, one could try to construct minimalistic cosmological
scenarios where both of them are attributed to the same source. This can be
performed by assuming a cosmic fluid with a Chaplygin gas equation of state
\cite{kamenshchik,Bilic:2001cg,Gorini:2002kf} or its generalization
\cite{Bento:2002ps,P83,P87,P85,Ali:2011sv,P86,P90,P91,P89}. In particular, in
this scenario the cosmic fluid behaves as pressureless at early times, thus
mimicking the matter epoch, and then its equation of state progressively
decreases, resulting to a cosmological constant at late times, and thus
mimicking the dark-energy domination. However, since the simple models are
not favored by the data \cite{P82}, many extensions appeared in the
literature, such as the modified Chaplygin gas (MCG) \cite{P92,P93,P94}, the
modified cosmic Chaplygin gas (MCCG) \cite{P96,P97}, and the extended Chaplygin
gas (ECG) \cite{Pourhassan:2014ika,Pourhassan:2014ikb,Pourhassan:2014ikc,Pourhassan:2014ikd,P103}.

Apart from the late-time acceleration and its previous matter epoch, a
successful cosmological scenario should be able to explain the initial phase
of the universe, that is the inflationary exponential expansion
\cite{Guth:1980zm,Linde:1981mu}, or alternatively the bounce realization
\cite{Novello:2008ra}. Unfortunately, the majority of the mentioned
models, although successful in obtaining the correct late-time behavior, are
unable to provide such a unified description of the whole thermal history of
the universe. Therefore, it is both interesting and worthy to try to
construct cosmological scenarios capable of providing such a unified
description.

In the present work, we construct a model where a cosmic fluid with an
extended Chaplygin gas equation-of-state interacts with a phantom field.
This scenario has many advantages, and the resulting cosmology proves to be
very interesting. In particular, we can obtain both inflationary and
bouncing solutions, thus being able to describe the initial universe phase
in both ways. Additionally, we can obtain the subsequent matter-dominated
phase and then the late-time acceleration, which completes the universe
history. We mention that the phantom field may lead the dark energy sector
to lie either in the quintessence, or in the phantom regime, or even exhibit
the phantom-divide crossing during the evolution, which is a great
advantage. Finally, the Chaplygin gas nature, provides as usual a unified
picture of the dark energy and dark matter sectors, offering a minimalistic
picture.\\
This paper organized as follows. In next section we introduce our model and write general
equations. In section 3 we obtain cosmological solutions for various situations. We find the early
and late time solutions, also discuss about inflationary and bouncing solutions. In section 4 we
study scalar perturbations and investigate stability of the model. Furthermore we use recent
observational data to fix some solutions. Finally in section 5 we give conclusions.

\section{The model}

In order to construct a model of a unification of the dark sector, which is
moreover capable of describing the phantom regime and alleviating the
coincidence problem we consider a cosmic fluid with an extended Chaplygin
gas equation of state, interacting with a phantom field. In particular, the
total action in the scenario at hand writes as
\begin{eqnarray}
S = \int d^{4}x \sqrt{-g} \left[\frac{1}{2\kappa^2} R +
\frac{1}{2}g^{\mu\nu}\partial_{\mu}\phi\partial_{\nu}\phi-V(\phi)
 \right]+S_{ecg}~,
 \label{action0}
\end{eqnarray}
where $\kappa^2=8\pi G$ is the gravitational constant, $R$ is the Ricci
scalar, $\phi$ is the phantom field with $V(\phi)$ its potential, and
$S_{ecg}$ is the action of the extended Chaplygin gas.
Throughout this work, we consider a spatially-flat Friedmann-Robertson-Walker
(FRW) metric of the form
\begin{equation}
\label{metric0}
ds^2= -dt^2+a^2(t)\,\delta_{ij} dx^i dx^j,
\end{equation}
where  $x^i$ are the co-moving spatial coordinates, $t$ is the cosmic time,
$a(t)$ is the scale factor and $H=\dot{a}/a$ is the Hubble function (dots
denote differentiation with respect to $t$). Concerning the phantom field,
and assuming homogeneity, its energy density and pressure are respectively
given as usual by
\begin{eqnarray}
\label{rhophant}
&&\rho_{\phi}=-\frac{1}{2}\dot{\phi}^{2}+V(\phi)\nonumber\\
&&p_{\phi}=-\frac{1}{2}\dot{\phi}^{2}-V(\phi).
\label{pphant}
\end{eqnarray}
Concerning the extended Chaplygin gas, in order to obtain a full generality
we consider an equation of state of the form \cite{P103,Pourhassan:2014ika}
\begin{equation}
\label{s4}
p_{ecg}=\sum_{n}A_{n}\rho_{ecg}^{n}-\frac{B}{\rho_{ecg}^{\alpha}},
\end{equation}
where $A_n$, $B$ and $n$ are constants. Note that in the case $n=1$ the above
expressions recovers the standard modified Chaplygin gas.

Variation of the action (\ref{action0}) with respect to the metric gives
rise to the two Friedmann equations, namely
\begin{equation}
\label{s10}
3H^{2}=\rho_{ecg}+\rho_{\phi},
\end{equation}
and
\begin{equation}\label{s11}
-2\dot{H}=\rho_{ecg}+\rho_{\phi}+p_{ecg}+p_{\phi},
\end{equation}
where we set $8\pi G=1$ for simplicity. In the scenario at hand,
$\rho_{ecg}$ incorporates both the dark matter and a part of the dark energy,
while $\rho_\phi$ is added to that part of the dark energy and completes the
dark energy sector. Finally, it proves convenient to define the deceleration
parameter
\begin{equation}
\label{decel}
q\equiv -1-\frac{\dot{H}}{H^2},
\end{equation}
which quantifies the accelerated nature of the expansion ($q<0$ corresponds
to acceleration).

Concerning the evolution equations of $\rho_\phi$  and $\rho_{ecg}$, the
total energy momentum tensor conservation allows for an interaction of the
form
\begin{equation}\label{s5}
\dot{\rho}_{ecg}+3H(\rho_{ecg}+p_{ecg})=-Q,
\end{equation}
and,
\begin{equation}\label{s6}
\dot{\rho}_{\phi}+3H(\rho_{\phi}+p_{\phi})=Q,
\end{equation}
where $Q$ is a general interaction term ($Q>0$ corresponds to energy transfer
from the extended Chaplygin gas to the phantom field and vice versa). There
are many proposed forms of $Q$ in the literatures, however a general
consensus is that it is reasonable to be proportional to some energy
density, since increased energy densities are expected to lead to increased
interactions. In the following we consider
\begin{equation}
\label{Q0}
Q=bH(\rho_{ecg}+\rho_\phi),
\end{equation}
with $b$ a constant, but other forms of interaction can be straightforwardly
incorporated.

\section{Background evolution}

First of all we assume that extended Chaplygin gas density is dominant, that
is we assume $\rho_{ecg}\gg\rho_{\phi}$, thus the phantom field is just a
small correction.

\subsection{Late-time behavior}

\subsubsection{The general case}

In the late time cosmology with the scale factor of the form $a(t)\propto
t^{\frac{2}{3(1+\omega)}}
$ one can obtain the
following Hubble expansion parameter,
\begin{equation}\label{s38}
H=\frac{2}{3(1+\omega)t},
\end{equation}
with the constant $\omega$. It yields to the following deceleration
parameter,
\begin{equation}\label{s39}
q=\frac{1}{2}+\frac{3}{2}\omega.
\end{equation}
It is obvious that $-1\leq\omega\leq-1/3$ gives $-1\leq q\leq0$.
Interaction term given by the Eq. (\ref{Q0}) simplified as follow,
\begin{equation}\label{s40}
Q=\frac{4b}{9(1+\omega)t^{3}}.
\end{equation}
Analyzing of this case verifies stability of our model at the late time and
confirm it.\\
There is also another interesting cosmological parameter so called jerk
parameter,
\begin{equation}\label{s41}
j=\frac{a^{2}}{{\dot{a}}^{3}}\frac{d^{3}a}{dt^{3}}=1+\frac{9}{2}
\omega(1+\omega).
\end{equation}
Some observational data such as $q=-1$, and $j=1$  \cite{P107} suggest
that $\omega=-1$. However this leat to a singularity at the Hubble expansion
parameter and interaction term $Q$. Therefore, we can choose $\omega=-0.5$ to
find $q=-0.25$, and $j=-0.125$ in agreement with the result of the Ref. \cite{P108}.
Therefore, we can use the late time scale factor as follow $a\propto t^{4/3}$.

\begin{figure}[h!]
 \begin{center}$
 \begin{array}{cccc}
\includegraphics[width=50 mm]{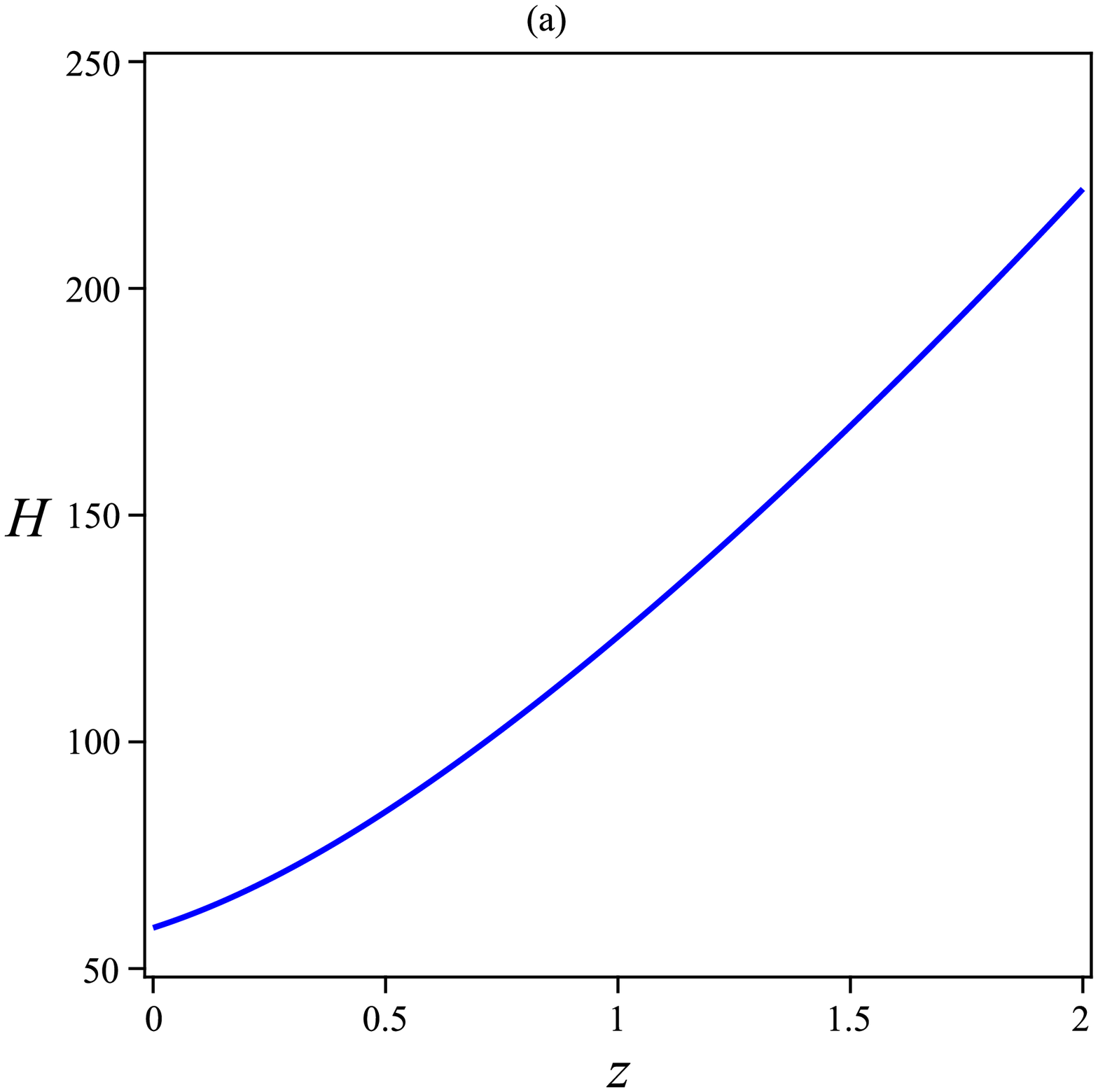}&\includegraphics[width=50
mm]{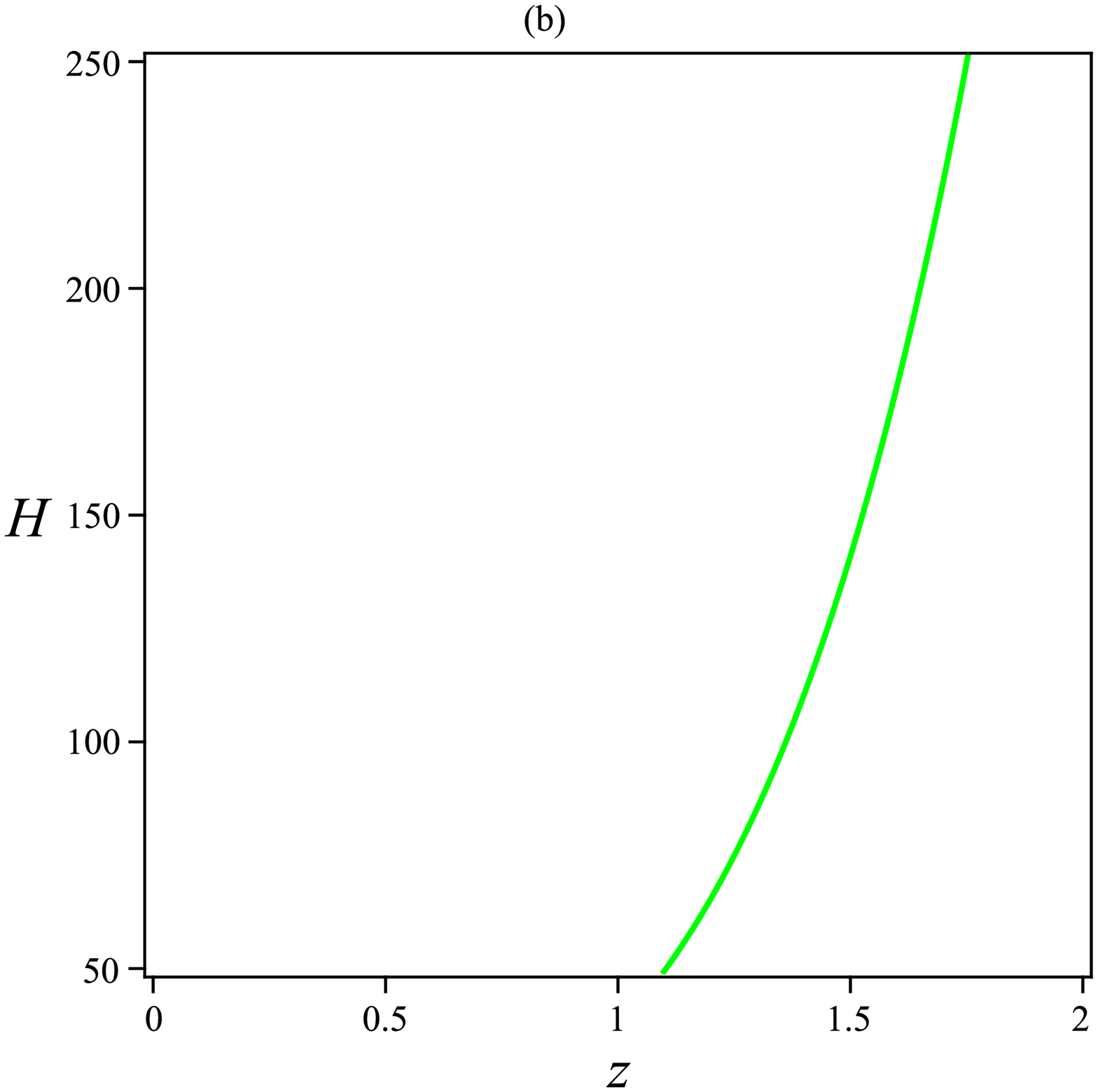}\\
\includegraphics[width=50 mm]{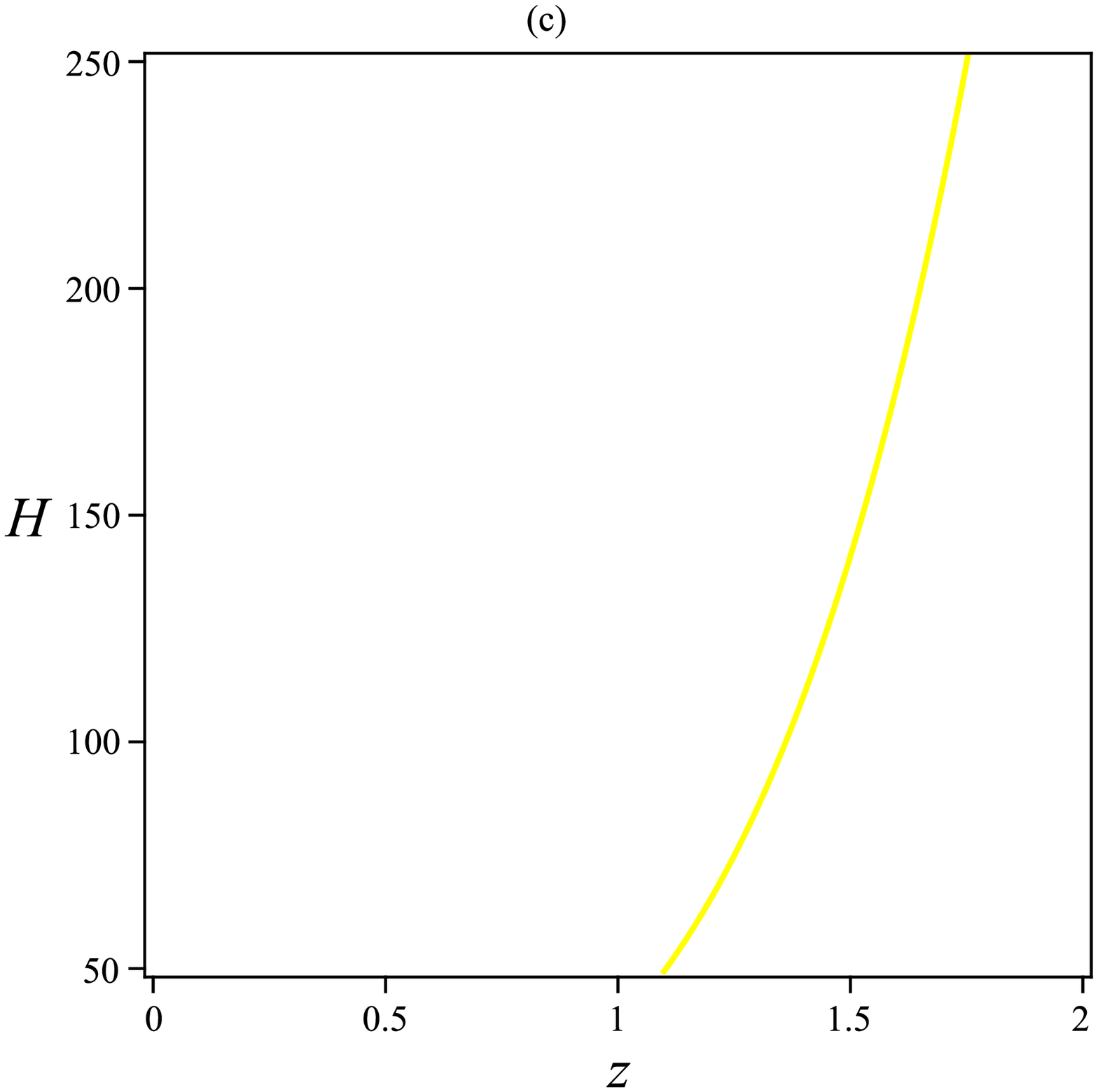}&\includegraphics[width=50
mm]{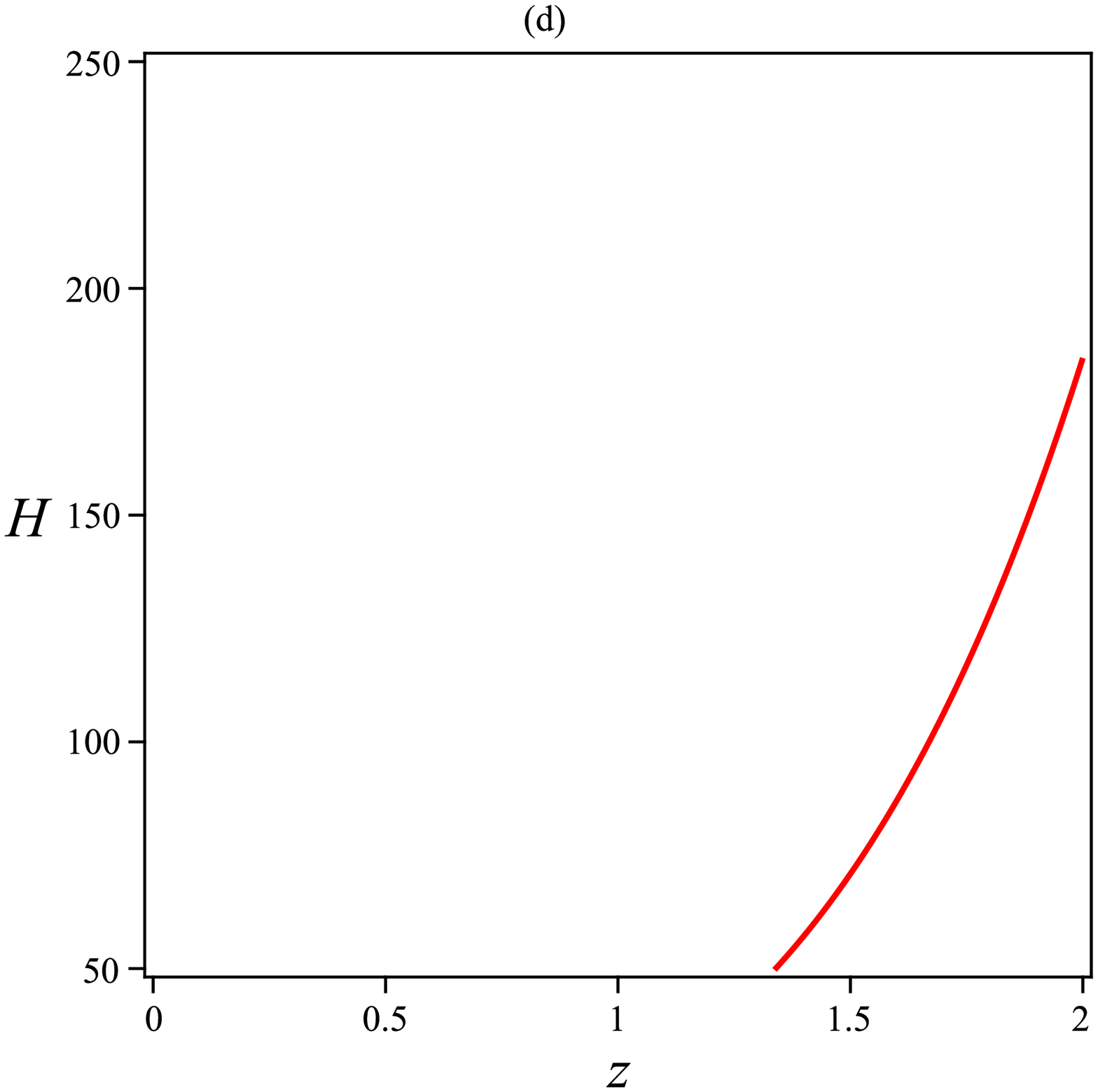}
 \end{array}$
 \end{center}
\caption{Hubble parameter in terms of redshift for $A=1/3$, $\alpha=1/2$. (a)
$n=1$, (b) $n=2$, (c) $n=3$, (d) $n=4$.}
 \label{fig:Hz}
\end{figure}

By using above observational data we can obtain Hubble expansion parameter in
terms of redshift which illustrated in the Fig. \ref{fig:Hz} for $n=1, 2, 3, 4$. It is
clear that the case of $n=1$ coincides with the observational data (for
example see the Ref.  \cite{P109}). We can see that cases of $n=2$ and $n=3$
increases value of the Hubble parameter at high redshift but decreases value
of the Hubble parameter at present ($z\sim0$). However, the case of $n=4$ may
be agree with the observational data at the high redshift. It tells that our
model under assumption $\rho_{ecg}\gg\rho_{\phi}$ may be valid at the high
redshift, means the early universe.

As we know the current value of the Hubble expansion parameter is a constant
in the range of $H_{0} = 65-75 km/s/Mpc$.
In that case, we have the following differential equations
corresponding to $\rho_{ecg}$ and $\rho_{\phi}$ respectively,
\begin{equation}\label{s50}
\dot{\rho}_{ecg}+3H_{0}\rho_{ecg}(1+\omega_{ecg})+\frac{9}{2}H_{0}^{2}=0,
\end{equation}
and,
\begin{equation}\label{s51}
\dot{\rho}_{\phi}-3H_{0}\rho_{\phi}(1+\omega_{\phi})+\frac{9}{2}H_{0}^{2}=0,
\end{equation}
where we should choose $-1.2\leq\omega_{\phi}\leq-1$ in agreement with
current observational data. Therefore, $\rho_{\phi}$ can extracted easily.
Hence we could decouple conserved equation to extract energy densities.

\subsubsection{The case $n=1$}
This is corresponding to MCG. We can obtain two possible solutions
depend on the
energy density of ECG.  If $\rho_{ecg}>1$, then we have the following energy density,
\begin{equation}\label{1}
\rho=-3H_{0}(5+\frac{1}{4(1+A)})+Ce^{-\frac{3}{10}H_{0}t}+\frac{\sqrt{16B(1+A)+9H_{0}^{2}
+8(1+A)e^{-
\frac{6H_{0}}{1+A}t}}}{4(1+A)},
\end{equation}
where $C$ is an integration constant. In order to obtain the above solution we used the
fact that $\tanh^{-1}(f(\rho))$ is not well defined for $\rho_{mcg}\geq1$.\\
On the other hand, for the case of $\rho_{mcg}\ll1$ we have the following energy density,
\begin{eqnarray}\label{2}
\rho=&-&3H_{0}(5+\frac{1}{4(1+A)})+Ce^{-\frac{3}{10}H_{0}t}\nonumber\\
&-&\frac{\sqrt{16B(1+A)+9H_{0}^{2}}}{4(1+A)}\tanh^{-1}\left[(1+A)\sqrt{16B(1+A)+9H_{0}^{2}
}t\right],
\end{eqnarray}
where $C$ is an integration constant. We should note that the above solution obtained for
$\alpha=1$. In both cases we can obtain $q=-1$ which is expected.

\subsection{Early-time behavior}

\subsubsection{The case $n=1$}

In the simplest case we consider the early universe cosmology where energy
density is very high and assume that $\alpha=0.5$ and $n=1$. Under these
assumptions the ECG behaves as barotropic fluid with linear EoS. In that case
we can obtain analytical solutions of the equations (\ref{s5}) and (\ref{s6}). The energy
density of the ECG can obtained as follow,
\begin{equation}\label{s24}
\rho_{ecg}=\left[\frac{\sqrt{6}}{2}(1+A+\frac{b}{3})t+C_{1}\right]^{-2},
\end{equation}
where $C_{1}$ is an integration constant and should be of order time scale at
the early universe to give high density. Therefore, we choose $C_{1}\ll1$ and
use relation (\ref{s24}) in the equation (\ref{s6}) to find,
\begin{equation}\label{s25}
\rho_{\phi}=\left[\frac{\sqrt{6}b}{((6A+b+3)(b+3)+9A^{2})t+C_{2}}\right]^{2},
\end{equation}
where $C_{2}$ in an integration constant and should be of order unity because
of our initial assumption $\rho_{ecg}\gg\rho_{\phi}$. In that case we can obtain Hubble
expansion
parameter as follow,
\begin{equation}\label{s26}
H\approx\frac{1}{3}\sqrt{\frac{4}{(1+A+\frac{b}{3})^{2}}t^{-2}+\frac{36b^{2}}
{\left(((6A+b+3)(b+3)+9A^{2})t+C_{2}\right)^{2}}}.
\end{equation}
where we neglected $C_{1}$ and used equations (\ref{s10}), (\ref{s24}) and (\ref{s25}).
So, using
the equation (\ref{decel}) we can investigate evolution of the deceleration parameter.
At the very early universe one can obtain the following expression,
\begin{equation}\label{s27}
q=-1+\mathcal{A}-\mathcal{B}t^{2}+\mathcal{O}(t^{3}),
\end{equation}
where constants $\mathcal{A}$ and $\mathcal{B}$ defined as,
\begin{eqnarray}\label{s28}
\mathcal{A}&\equiv&\frac{3A+b+3}{2},\nonumber\\
\mathcal{B}&\equiv&\frac{3b^{2}(3A+b+3)^{3}}{4C_{2}^{2}}.
\end{eqnarray}
It is clear that $-1+\mathcal{A}$ is positive for positive values of $b$ and
$A$, hence the value of $q$ is positive at the early universe which is
expected.

\subsubsection{The case $n=2$}

\begin{figure}[h!]
 \begin{center}$
 \begin{array}{cccc}
\includegraphics[width=60 mm]{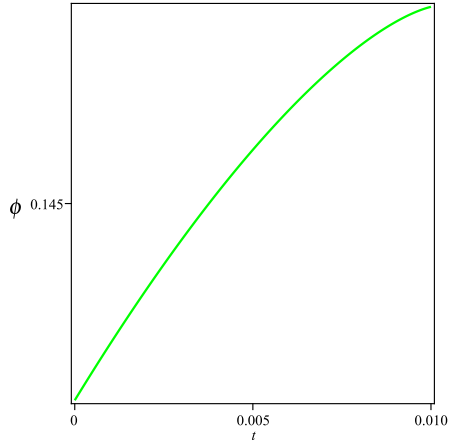}
 \end{array}$
 \end{center}
\caption{Scalar field $\phi$ in terms of $t$ for $A=1/3$, $C_{3}=1$,
$C_{4}=1$, $\alpha=1/2$, $b=1$, $\omega_{\phi}=-1$ and $n=2$.}
 \label{phi1}
\end{figure}

Again, we consider the early universe cosmology where energy density is very
high and assume that $\alpha=0.5$ and $n=2$. Under these assumptions the ECG
behaves as barotropic fluid with quadratic EoS. In that case we can obtain
analytical solutions of the equations (\ref{s5}) and (\ref{s6}). First, we can obtain the
following expression from the equation (\ref{s5}),
\begin{equation}\label{s29}
t-\frac{3\sqrt{2A}\tan^{-1}\left(\sqrt{\frac{3A\rho_{ecg}}{3A+b+3}}\right)}{
(3A+b+3)^{\frac{3}{2}}}-\frac{\sqrt{6}}{(3A+b+3)\sqrt{\rho_{ecg}}}+C_{3}=0,
\end{equation}
where $C_{3}$ is an integration constant. At the initial time with the high
density we can obtain,
\begin{equation}\label{s30}
\rho_{ecg}\approx \frac{3A+b+3}{3A}\tan^{2}{\tilde{t}},
\end{equation}
where $\tilde{t}\equiv \frac{\sqrt{2A}(3A+b+3)^{\frac{3}{2}}(t+C_{3})}{6A}$
is used. In that case one can obtain,
\begin{equation}\label{s31}
\rho_{\phi}\approx
\frac{\sqrt{6}b(3A+b+3)(\tilde{t}-\tan{\tilde{t}})+C_{4}}{324A^{2}},
\end{equation}
where $C_{4}$ is another integration constant.\\
Easily we can check that the Hubble expansion parameter is decreasing
function at the initial times. Also, analysis of the deceleration parameter
tells that the universe may begin with the negative $q$, if value of $b$ be
smaller than one, then transfer to the universe with positive $q$ and again
return to the deceleration phase with the negative $q$. On the other hand,
with $b>1$ we have initial universe in acceleration phase which transferred
to the deceleration phase at the late time.\\
Behavior of the interaction term $Q$ is interesting, it has periodic like
feature during time. It begin with a high value at initial time and decreased
to a small value and again grows up to a high value, and repeat this
procedure. However, it is corresponding to the early universe.\\
It is useful to investigate behavior of the scalar field. We can use some assumption about
the
scalar potential $V(\phi)$
from the previous works such as the PNGB model \cite{P105} with a
phantom potential given by,
\begin{equation}\label{s32}
V(\phi)=\rho_{\phi0}- M^{4}(1-\cos{\frac{\phi}{f}}),
\end{equation}
where $M$ and $f$ are constants and $\rho_{\phi0}$ is present value of
density which can considered as a constant. There is also other model with a
local phantom potential including the Gaussian potential \cite{Dutta:2009yb},
\begin{equation}\label{s33}
V(\phi)=\rho_{\phi0}- M^{4}(1-e^{\frac{\phi^{2}}{\sigma^{2}}}),
\end{equation}
where $\sigma$ is a constant. Moreover the quadratic phantom potential may be
written as \cite{Dutta:2009yb},
\begin{equation}\label{s34}
V(\phi)=\rho_{\phi0}-V_{2}\phi^{2},
\end{equation}
where $V_{2}$ is a constant. All of the above models for the infinitesimal
values of the scalar fields behave as the following,
\begin{equation}\label{s35}
V(\phi\ll1)=a_{1}+a_{2}\phi^{2}+\mathcal{O}(\phi^{3}),
\end{equation}
where $a_{1}$ and $a_{2}$ are arbitrary constants. In that case we can obtain
typical behavior of the scalar field in terms of time corresponding to our
model as illustrated in Fig. \ref{phi1}. We can see that the scalar field is
increasing function of time, and its evolution in the early universe is faster than that in the late time.\\

\subsubsection{The case $n>2$}

\begin{figure}[h!]
 \begin{center}$
 \begin{array}{cccc}
\includegraphics[width=60 mm]{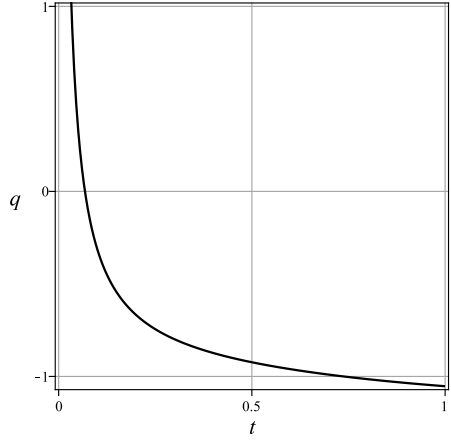}
 \end{array}$
 \end{center}
\caption{Deceleration parameter $q$ in terms of $t$ for the case of larger
$n$.}
 \label{Q3}
\end{figure}

For the larger $n$, at high density regions, we can consider only the last
term of the expansion in EoS given by the Eq. (\ref{s4}). In that case the solution of the
equations (\ref{s5}) and (\ref{s6}) obtained easily as follow,
\begin{equation}\label{s36}
\rho_{ecg}=\left[\frac{2}{\sqrt{6}A(2n-1)t+2C_{5}}\right]^{\frac{2}{2n-1}},
\end{equation}
and,
\begin{equation}\label{s37}
\rho_{\phi}=\left[\frac{2}{\sqrt{6}A(2n-1)t+2C_{5}}\right]^{\frac{4}{2n-1}}\times\frac{A^{2}b^{2}(2n-1)^2}{(A(2n-3)+C_{6})^2}t^{2},
\end{equation}
where $C_{5}$ and $C_{6}$ are integration constants and should be considered
as $C_{5}\ll1$ and $C_{6}\approx1$ to satisfy $\rho_{ecg}\gg\rho_{\phi}$
condition, so we can neglect $C_{5}$.\\
We find that the Hubble expansion parameter is decreasing function of time
which decreased also by increasing $n$. Also, we can investigate evolution of
the deceleration parameter and see that it begin with positive value and
yields to -1 at the late time (see Fig. \ref{Q3}), which shows deceleration to acceleration phase transition.

\subsection{Special solution}
\begin{figure}[h!]
 \begin{center}$
 \begin{array}{cccc}
\includegraphics[width=60 mm]{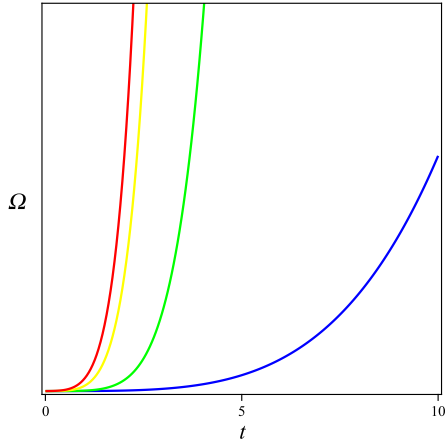}
 \end{array}$
 \end{center}
\caption{Typical behavior of $\Omega_{ecg}$ and $\Omega_{\phi}$ in terms of cosmic time.
$n=1$ blue,
 $n=2$ green, $n=3$ yellow, $n=4$ red.}
 \label{coincident}
\end{figure}

In the previous subsections we restricted ourself on the case of higher ECG energy density
than
scalar field energy density. This assumption simplified our equations and help us to
obtain some
analytical expressions. Now, we try to study more general case without any condition on
the energy
densities. In that case, from the conservation energy equations, one can obtain,
\begin{equation}\label{s42}
\dot{\rho}_{ecg}+3H(1+\omega_{ecg})\rho_{ecg}+3bH^{3}=0,
\end{equation}
and,
\begin{equation}\label{s43}
\dot{\rho}_{\phi}+3H(1+\omega_{\phi})\rho_{\phi}-3bH^{3}=0.
\end{equation}
We can consider above relations as cubic equation for $H$ and solve them to obtain,
\begin{eqnarray}\label{s44}
H&=&\frac{X^{\frac{1}{3}}}{6b}-\frac{2(1+\omega_{ecg})\rho_{ecg}}{X^{\frac{1}{3}}}
\nonumber\\
&=&\frac{Y^{\frac{1}{3}}}{6b}+\frac{2(1+\omega_{\phi})\rho_{\phi}}{Y^{\frac{1}{3}}},
\end{eqnarray}
where we defined,
\begin{eqnarray}\label{s444}
X&\equiv&\left(-36\dot{\rho}_{ecg}+12\sqrt{3}\sqrt{\frac{3b\dot{\rho}_{ecg}^{2}+4
\rho_{ecg}^{3}(1+3\omega_{ecg}(1+\omega_{ecg})+\omega_{ecg}^{3})}{b}}\right)b^{2},
\nonumber\\
Y&\equiv&\left(36\dot{\rho}_{\phi}+12\sqrt{3}\sqrt{\frac{3b\dot{\rho}_{\phi}^{2}-4\rho_{
\phi}^{3}(1+3\omega_{\phi}(1+\omega_{\phi})+\omega_{\phi}^{3})}{b}}\right)b^{2}.
\end{eqnarray}
Therefore, having $\rho_{ecg}$, can gives us Hubble parameter as well as $p_{ecg}$. Then,
from the
equation (\ref{s10}) one can obtain $\rho_{\phi}$ and use it in the equation (\ref{s11})
to find $p_
{\phi}$. Moreover, we can use the equations (\ref{rhophant}) to obtain scalar field and
scalar
potential. So, we can use some ansatz for $\rho_{ecg}$ to have all cosmological
parameter
and compare them with observational data.\\
Here, we assume that the extended Chaplygin gas has the following form,
\begin{equation}\label{ansatz}
\rho_{ecg}=\sqrt{a_{1}+\frac{a_{2}}{t^{m}}},
\end{equation}
where $a_{1}$, $a_{2}$ and $m$ are free parameters which will be fixed using recent
observational
data. The ansatz (\ref{ansatz}) tells that $\rho_{ecg}\rightarrow\infty$ at the early
universe and $\rho_{ecg}\rightarrow\sqrt{a_{1}}$ at the late time. These are agree with
the fact that the energy density should be decreasing function of time in an expanded universe. We expected that
the parameter $a_{1}$ be small to have small energy density at the late time. This yields to a
Hubble parameter (as well as $\dot{\phi}$) as decreasing function of time which is expected. Also, we can
find that the deceleration parameter yields to -1 at the late time, with possibility to
have
acceleration to deceleration phase transition. All of them depend on choosing appropriate
parameters. We will fix them using recent observational data in section 4. Interestingly, we can
analyze,
\begin{equation}\label{Omeg1}
\Omega_{ecg}=\frac{\rho_{ecg}}{3H^{2}}a^{2},
\end{equation}
and,
\begin{equation}\label{Omeg2}
\Omega_{\phi}=\frac{\rho_{\phi}}{3H^{2}}a^{2},
\end{equation}
to see the exact equality $\Omega_{ecg}=\Omega_{\phi}$. It is illustrated in the Fig.
\ref{coincident}. For various $n$ the curves of $\Omega_{ecg}$ and $\Omega_{\phi}$ have
overlap and recovers each other exactly. It will be happen even for $n=1$, therefore interaction term
may be
cause of this event.

\subsection{Inflationary solutions}

\begin{figure}[h!]
 \begin{center}$
 \begin{array}{cccc}
\includegraphics[width=60 mm]{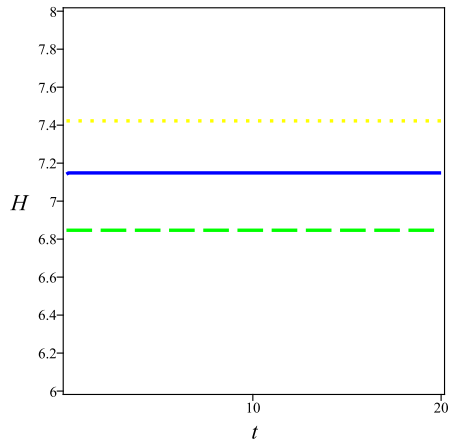}
 \end{array}$
 \end{center}
\caption{Hubble parameter versus time for $\alpha=1$, $B=3.4$,
$A_{n}=\frac{n}{3}$ and $C=85$. $n=1$ (solid line), $n=2$ (dashed line),
$n=3$ (dotted line).}
 \label{fig:z}
\end{figure}

Finding $H$ approximately as a constant, we can investigate inflationary
solutions in the model of interacting extended Chaplygin gas with phantom
field. Recently, an inflationary model of modified Chaplygin gas with quintessence field has been studied \cite{Emre}. The present work may be extension of that paper to the case of ECG in the context of phantom cosmology. Therefore, there are three main differences between this work and Ref. \cite{Emre}:\\
The first: extension of MCG to ECG.\\
The second: exchange of quintessence field to phantom field.\\
The third: consideration of interaction term.\\
In order to obtain such solutions we combine equations (\ref{s5}), (\ref{s6}) and
(\ref{s10}) to find the following differential equation,
\begin{equation}\label{s53}
\dot{H}-\frac{\dot{\rho}_{\phi}}{6H}+\frac{1}{2}H^{2}(3+b)-\frac{\rho_{\phi}}
{2}+\frac{1}{2}\sum{A_{n}
[\frac{3}{2}H^{2}-\rho_{\phi}]^{n}}-\frac{B}{[\frac{3}{2}H^{2}-\rho_{\phi}]^{
\alpha}}=0.
\end{equation}
The simplest inflationary solutions obtained under assumption of constant
$\rho_{\phi}$. Fig. \ref{fig:z} shows that Hubble parameter obtained as a constant if we
assume $\rho_{\phi}=C$, where $C$ is an arbitrary constant. We exam our model
for $n=1, 2, 3$ and found that the Hubble expansion parameter is a constant
during cosmic time. Therefore, We can find ECG energy density as the
following constant,
\begin{equation}\label{s54}
\rho_{ecg}=\frac{3}{2}H^{2}-C.
\end{equation}

\subsection{Bouncing solutions}

\begin{figure}[h!]
 \begin{center}$
 \begin{array}{cccc}
\includegraphics[width=60 mm]{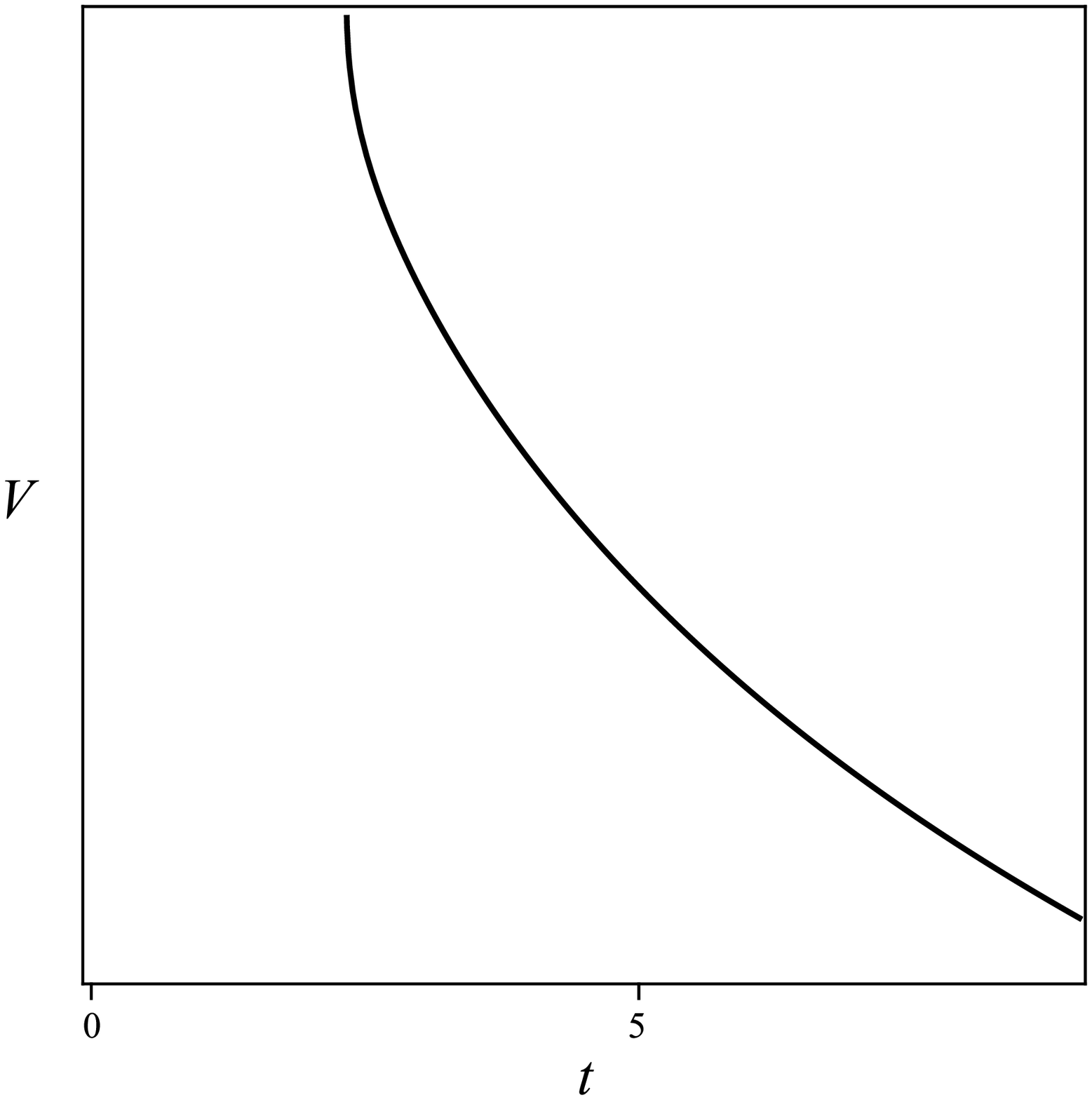}\includegraphics[width=60 mm]{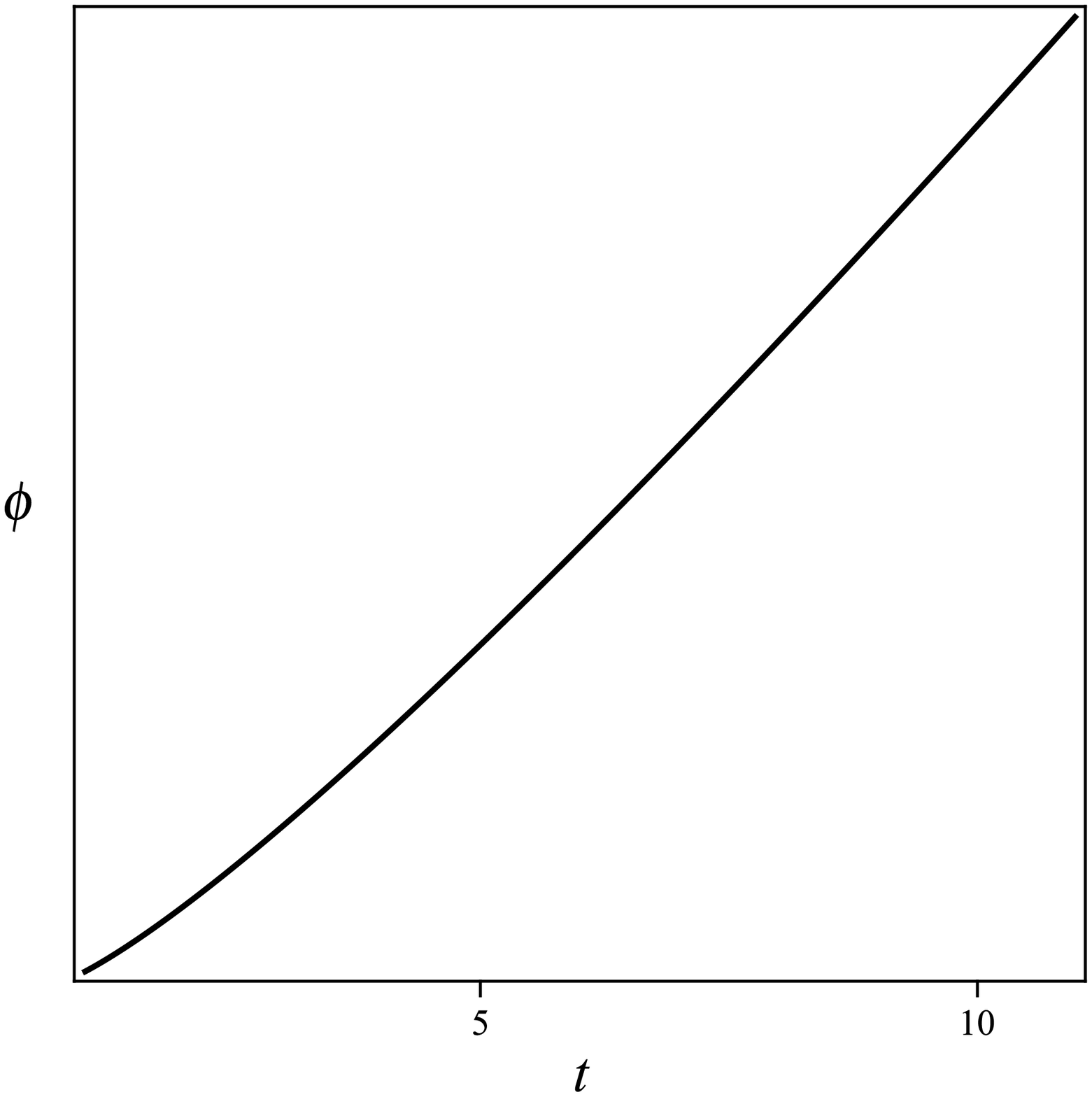}
 \end{array}$
 \end{center}
\caption{Typical behavior of scalar potential and scalar field in terms of
cosmic time for $\alpha=0.5$.}
 \label{fig:V}
\end{figure}

An interesting solution of the singularity problem of the standard cosmology
known as bouncing solution
\cite{Novello:2008ra,Shtanov:2002mb,Creminelli:2007aq,Cai:2011tc,Cai:2012ag,
Qiu:2013eoa,Quintin:2014oea}. The scale factor is decreasing in the contracting
phase, and increasing in the expanding phase. It means that $H<0$ transits to
$H>0$. We can reconstruct scalar potential as logarithmic function with
behavior represented in the left panel of the Fig. \ref{fig:V}. Then, we can
obtain behavior of the scalar field with time (see right panel of the Fig.
\ref{fig:V}).

\begin{figure}[h!]
 \begin{center}$
 \begin{array}{cccc}
\includegraphics[width=60 mm]{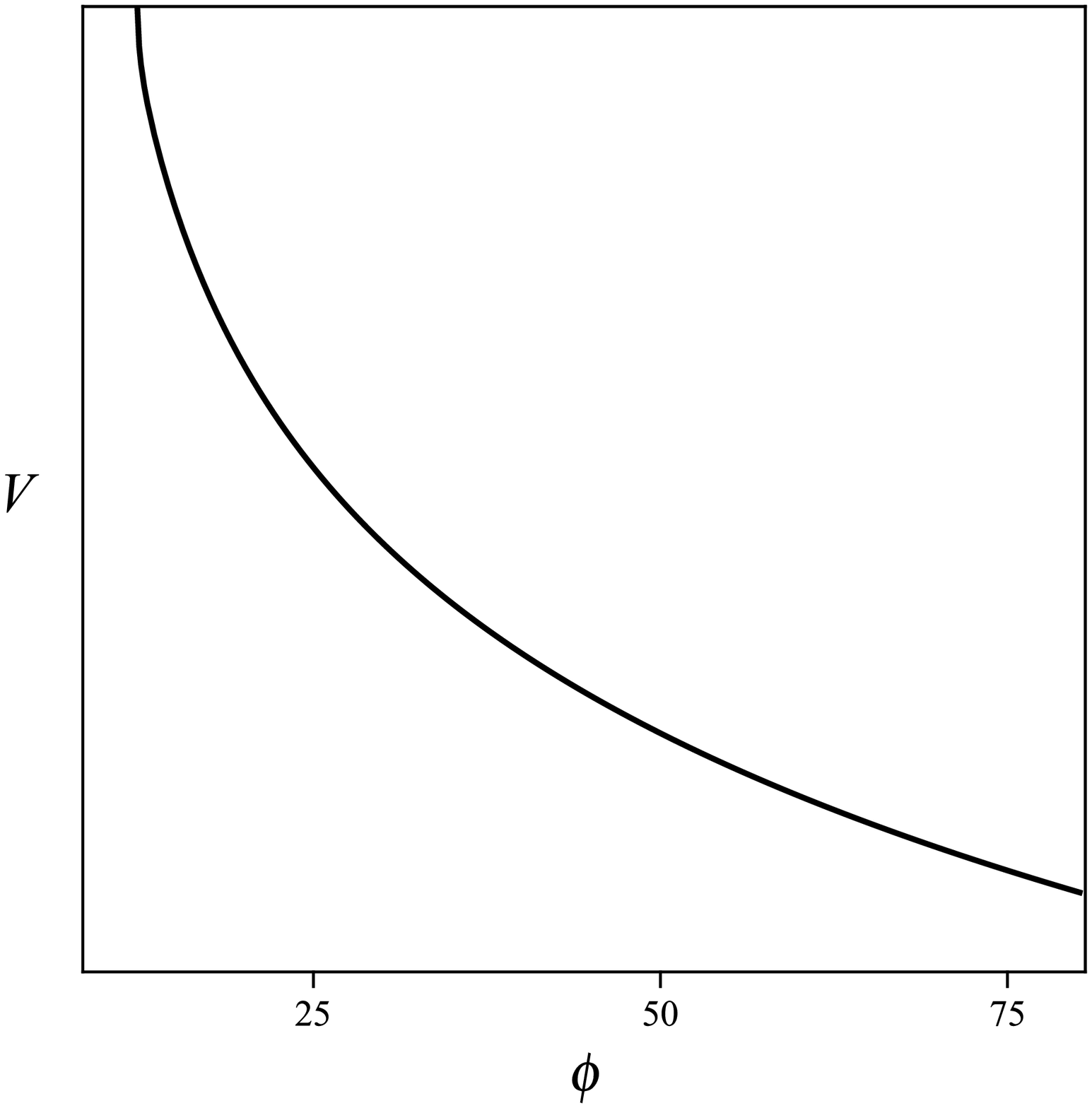}
 \end{array}$
 \end{center}
\caption{Typical behavior of scalar potential in terms of scalar field for $\alpha=0.5$.}
 \label{fig:VU}
\end{figure}

Therefore, we can obtain behavior of the scalar potential in terms of the
scalar field in the Fig. \ref{fig:VU}. We find that the scalar potential is
combination of some logarithmic functions and totally it is decreasing
function of scalar field.

\begin{figure}[h!]
 \begin{center}$
 \begin{array}{cccc}
\includegraphics[width=60 mm]{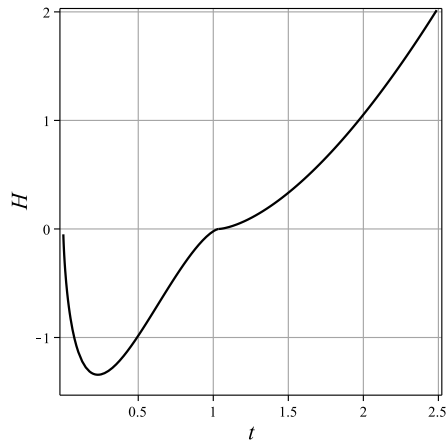}
 \end{array}$
 \end{center}
\caption{Hubble parameter versus time for $d_{1}=5$ and $d_{2}=0.965$.}
 \label{bounce4}
\end{figure}

The scalar potential has complicated and long expression with logarithmic
nature. Motivated by this behavior we can reconstruct scalar potential with
the simple logarithmic form,
\begin{equation}\label{s555}
V(\phi)=d_{1}\ln{[d_{2}(\phi-\phi_{0})]},
\end{equation}
where $d_{1}$ and $d_{2}$ are arbitrary constants. This is very simplified
potential but gives the similar results. We can use it to obtain
$\rho_{\phi}$ and therefore, $\dot{\phi}$. Then, we can find Hubble parameter
via the relation,
\begin{equation}\label{s5555}
H=\frac{V}{V^{\prime}}\dot{\phi},
\end{equation}
where $V^{\prime}\equiv\frac{dV}{d\phi}$. It yields to the Hubble parameter
represented by the Fig. \ref{bounce4}.
We can see that the Hubble parameter is negative for $t<1$ with
$\omega_{\phi}>-1$ and it is positive for $t>1$ with $\omega_{\phi}<-1$ (see
Fig. \ref{bounce4}).

\begin{figure}[h!]
 \begin{center}$
 \begin{array}{cccc}
\includegraphics[width=60 mm]{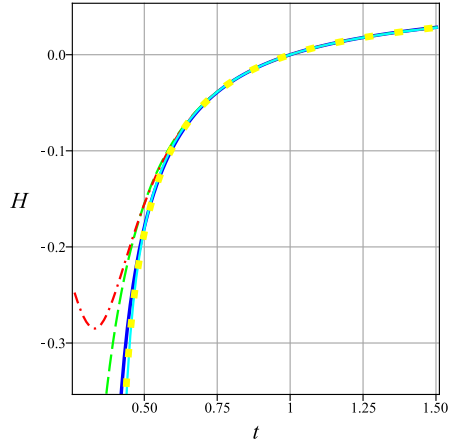}
 \end{array}$
 \end{center}
\caption{Hubble parameter versus time for $\alpha=1$, $B=3.4$,
$A_{n}=\frac{1}{3}$, $b_{1}=1$ and $b_{2}=1$. $n=1$ (solid blue line), $n=2$
(dashed green line), $n=3$ (dotted yellow line), $n=4$ (dash dotted red
line), $n=5$
(solid cyan line).}
 \label{fig:bounce}
\end{figure}

It yields us to the following general solution for the
ECG energy density,
\begin{equation}\label{s55}
\rho_{ecg}=b_{1}\ln{b_{2}t},
\end{equation}
where $b_{1}$ and $b_{2}$ are constant parameters related with $d_{1}$ and
$d_{2}$. In that case,
using the equations (\ref{s5}), (\ref{s6}) we can obtain behavior of the
Hubble expansion
parameter (see Fig. \ref{fig:bounce}). It means that by using logarithmic
behavior of the scalar potential and
ECG energy density the scale factor decreased first, corresponding to
negative
value of the Hubble expansion parameter, and increased after some time
corresponding to positive value of the Hubble expansion parameter.\\

We can see that in the cases of $n=1, 2, 3$, the Hubble parameter is
completely increasing function of time which yield to a constant at the late
time. The transition form $H<0$ to $H>0$ happen at $t=1$ for selected values
of parameters. In the case of $n=4$ the Hubble parameter decreased first to a
minimum negative value, then increases to positive value. The phase
transition time is similar to the previous cases. Cases with higher $n$ yield
to initially positive $H$ which suddenly take negative value after the early
universe and behave as previous cases after the initial time. It is obvious
that, at the late time there is no differences between various $n$. The fact
is that, importance of the first term in the right hand side of the equation
(\ref{s4})
is in the early universe. After some time, the second term is dominant.\\
In that case the scalar field energy density is given by,
\begin{equation}\label{s56}
\rho_{\phi}=3H^{2}-b_{1}\ln{b_{2}t},
\end{equation}
where,
\begin{eqnarray}\label{s57}
H=\frac{1}{3}b_{1}\ln{b_{2}t}\left[1+nA(b_{1}\ln{b_{2}t})^{n-1}-\frac{B}{(b_{
1}\ln{b_{2}t})^{1+\alpha}}\right]\times\nonumber\\
\left[-1+\sqrt{1-\frac{2}{b_{1}\ln{(b_{2}t)^{2}}\left((1+nAb_{1}\ln{b_{2}t})^
{n-1}-\frac{B}{(b_{1}\ln{b_{2}t})^{1+\alpha}}\right)^{2}}}\right],
\end{eqnarray}
and we assumed $A_{n}=A$.

\section{Perturbations}

\subsection{Scalar perturbations}
\begin{figure}[h!]
 \begin{center}$
 \begin{array}{cccc}
\includegraphics[width=50 mm]{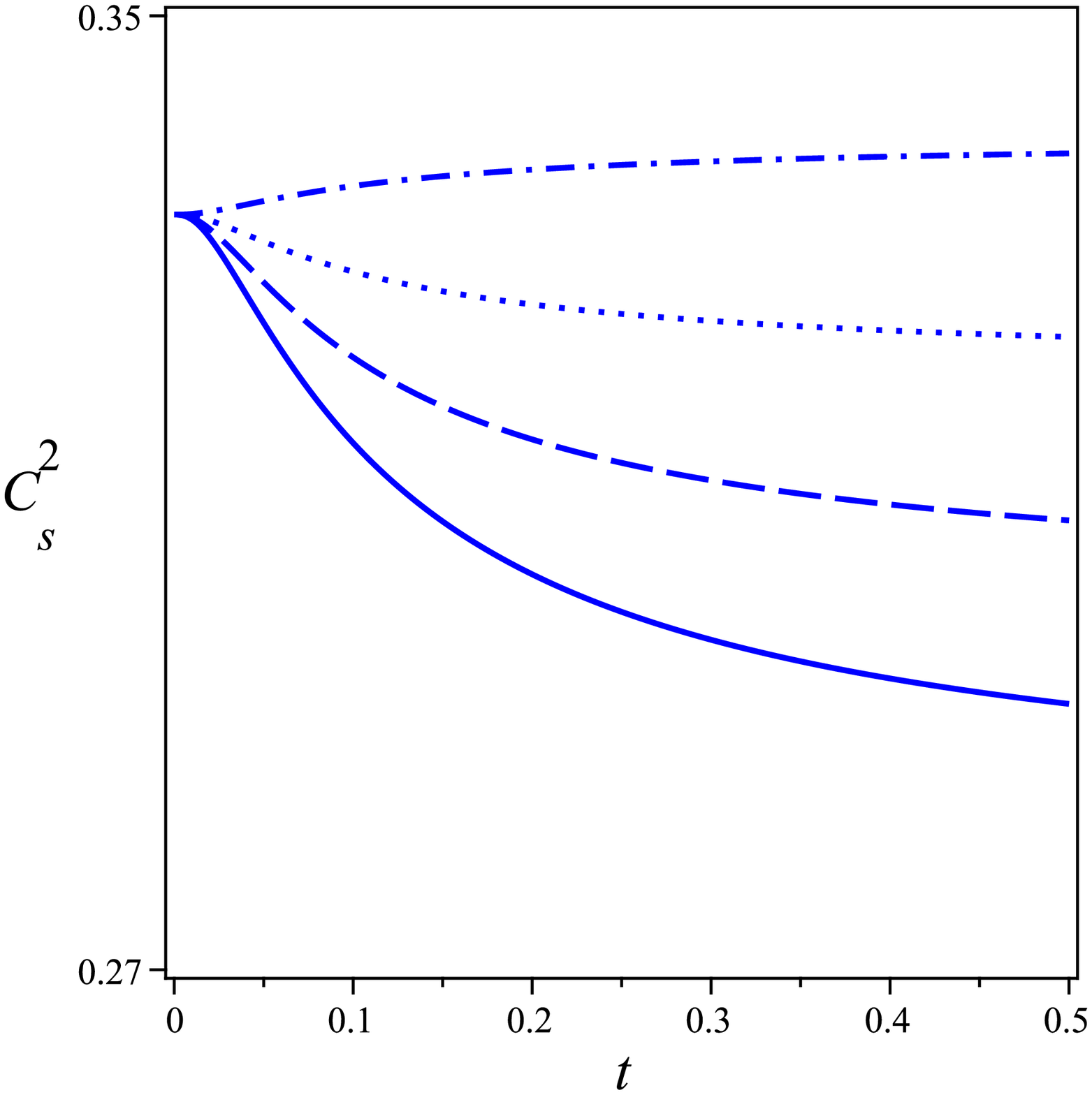}&
\includegraphics[width=50 mm]{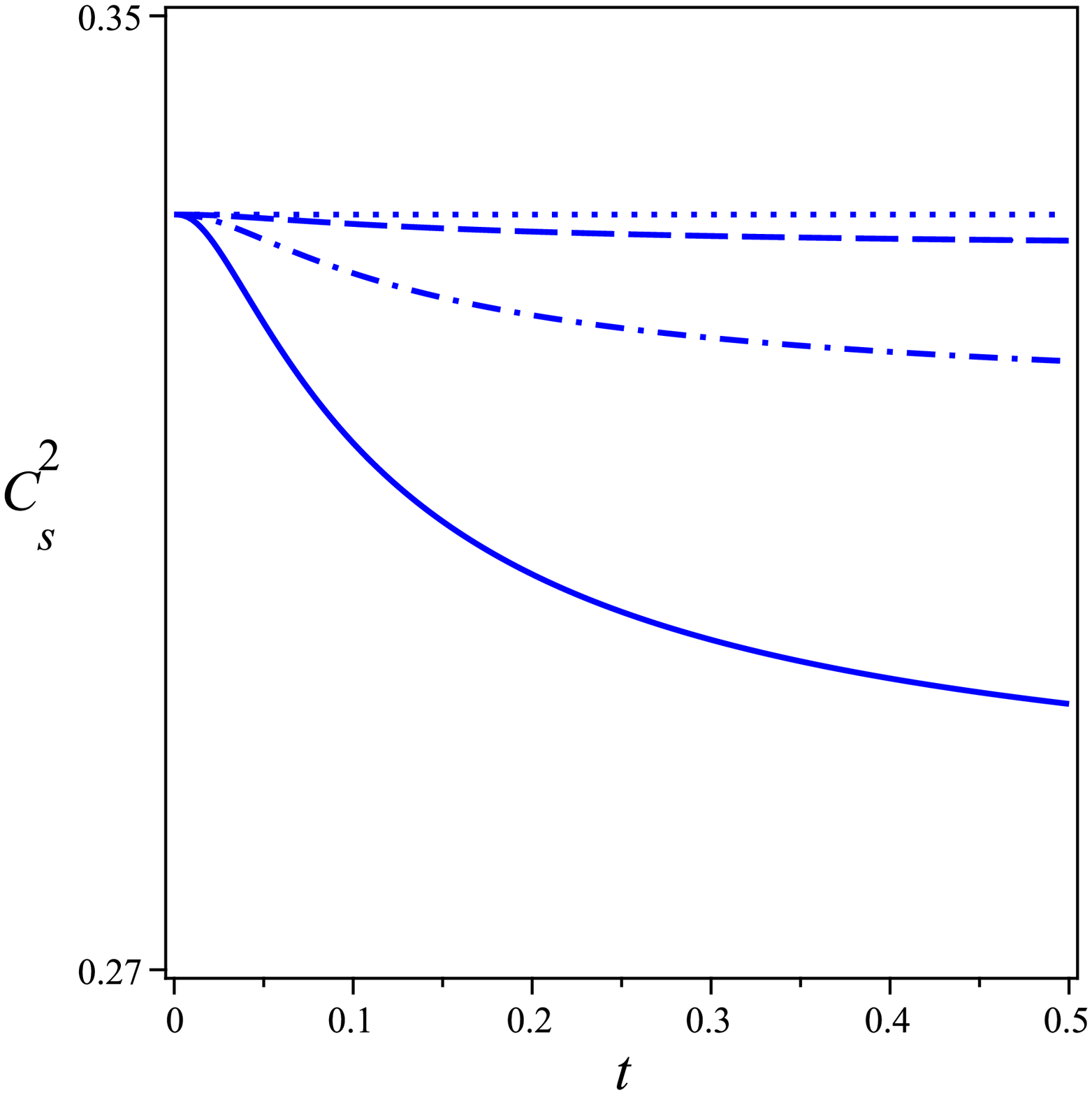}
 \end{array}$
 \end{center}
\caption{Squared sound speed in terms of $t$ for $A=1/3$, $C_{1}\ll1$,
$C_{2}=1$, $\alpha=1/2$ and $n=1$. Left: $b=1$, $\omega_{\phi}=0.5$ (dash
dotted line), $\omega_{\phi}=0$ (dotted line), $\omega_{\phi}=-0.5$ (dashed
line), $\omega_{\phi}=-1$ (solid line). Right: $\omega=-1$, $b=0.5$ (dash
dotted line), $b=0$ (dotted line), $b=0.2$ (dashed line), $b=1$ (solid
line).}
 \label{stab1}
\end{figure}

In this section we give density perturbation analysis of our model. Already,
density perturbation of a universe dominated by Chaplygin gas was studied by
the Ref. \cite{P110}. Now, we use their results to write the following
perturbation equation,
\begin{equation}\label{s52}
\ddot{\delta}+H[2-3(2\omega-C_{s}^{2})]\dot{\delta}-\frac{3}{2}H^{2}(1-6C_{s}
^{2}-3\omega^{2}+8\omega)\delta=-k^{2}\frac{C_{s}^{2}}{a^{2}}\delta,
\end{equation}
where $\delta$ is a density fluctuation, $k$ is the wave number of the
Fourier mode of the perturbation, and
\begin{equation}\label{s22}
C_{s}^{2}=\frac{\dot{p}}{\dot{\rho}},
\end{equation}
is squared sound speed.\\
Analytical solutions help us to investigate density
perturbation and stability analysis of the model. We can investigate
stability of the model by using squared sound speed.
If $C_{s}^{2}\geq0$ then, the model is stable. We sould note that it is different with the gravitational instabilities allow for structure formation as observed in the sky.\\
In the simplest case, $n=1$ we can find that
our model is stable at the initial time (see Fig. \ref{stab1}). We can see from Fig.
\ref{stab1} that the case of non-interacting ($b=0$) yields to constant sound speed. By
increasing strength of interaction, the sound speed decreased and yields to a
constant at the late time. So, the model is completely stable for $n=1$.\\

\begin{figure}[h!]
 \begin{center}$
 \begin{array}{cccc}
\includegraphics[width=60 mm]{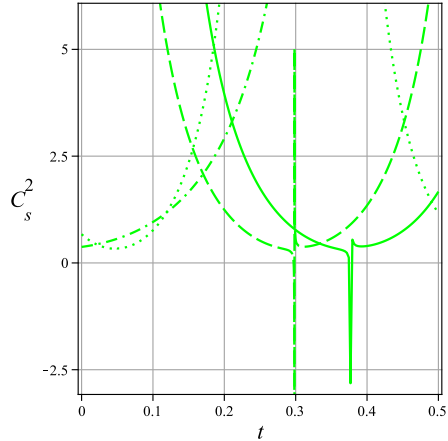}
 \end{array}$
 \end{center}
\caption{Squared sound speed in terms of $t$ for $A=1/3$, $C_{3}=1$,
$C_{4}=1$, $\alpha=1/2$, $\omega_{\phi}=-1$ and $n=2$. $b=2$ (dotted line),
$b=1.2$ (dashed line), $b=1$ (solid line), $b=0$ (dash dotted line).}
 \label{stab2}
\end{figure}

In the case of $n=2$ of the early universe we can obtain stability condition. Fig.
\ref{stab2}
shows that there are some instabilities
depend on value of the interaction strength. However, instabilities vanished
soon and our model will be stable. Opposite of the previous model with $n=1$,
the case of $b=0$ in present model with $n=2$ don't lead to the constant
sound speed.\\

\begin{figure}[h!]
 \begin{center}$
 \begin{array}{cccc}
\includegraphics[width=50 mm]{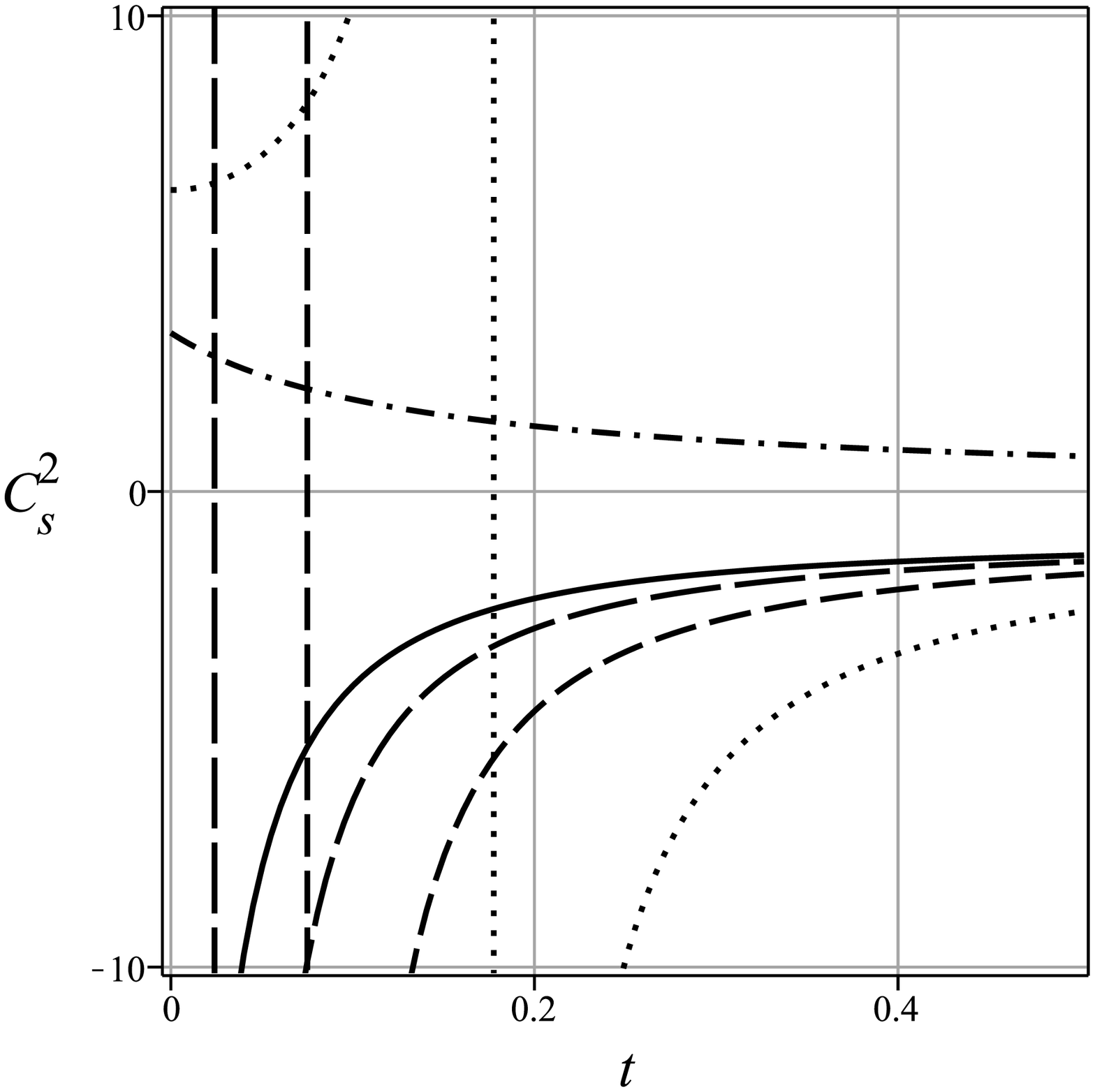}&
\includegraphics[width=50 mm]{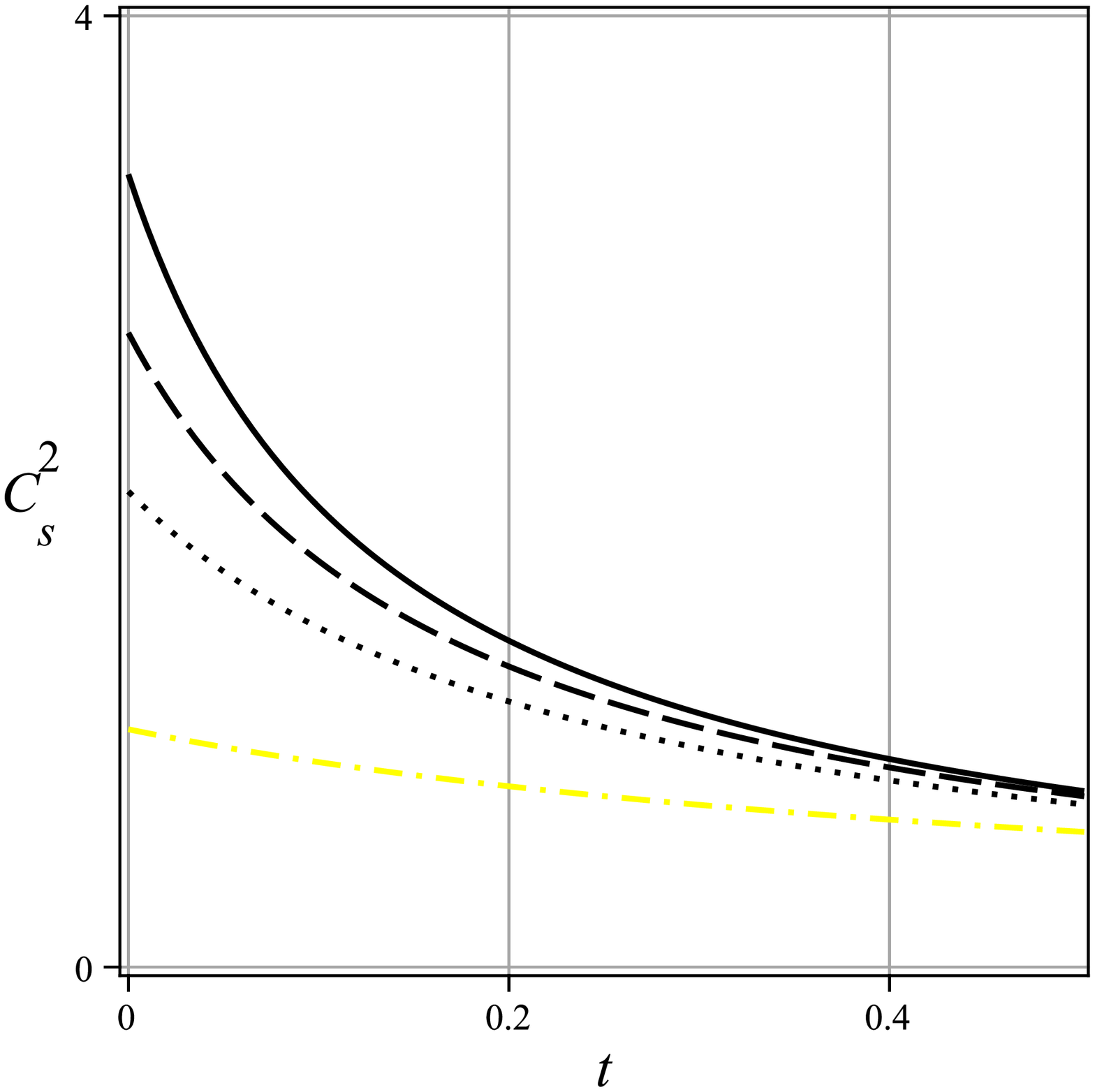}
 \end{array}$
 \end{center}
\caption{Squared sound speed in terms of $t$ for $A=1/3$, $C_{5}\ll1$,
$C_{6}=1$, $\alpha=1/2$ and $\omega=-1$ in the case of $n>2$. Left:
$n=10$, $b=0$ (dash dotted line), $b=0.4$ (dotted line), $b=0.6$ (dashed
line), $b=0.8$ (long dashed line), $b=1$ (solid line). Right: $b=0$, $n=3$
(dash dotted line), $n=6$ (dotted line), $n=8$ (dashed line), $n=10$ (solid
line).}
 \label{stab3}
\end{figure}

For the case of $n>2$ we can see behavior of squared
sound speed in Fig. \ref{stab3}. We can see that presence of interaction makes some
instabilities at the initial times. So, in the case of $b=0$ the model is
completely stable at the early universe. However, for the small values of $b$
($b\ll1$) we have stable model at the initial stage.

\begin{figure}[h!]
 \begin{center}$
 \begin{array}{cccc}
\includegraphics[width=60 mm]{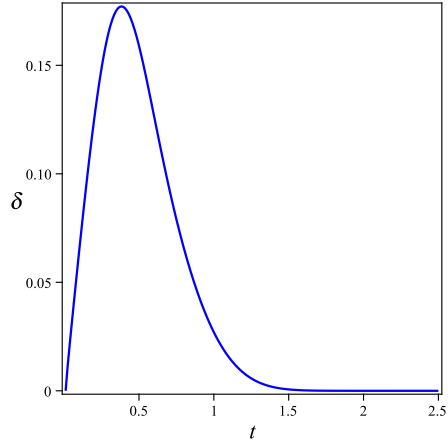}
 \end{array}$
 \end{center}
\caption{Time evolution of $\delta$ for the simplest case:
$\rho_{ecg}\gg\rho_{\phi}$ and $n=1$ at the early universe.}
 \label{fig:del1}
\end{figure}

\begin{figure}[h!]
 \begin{center}$
 \begin{array}{cccc}
\includegraphics[width=60 mm]{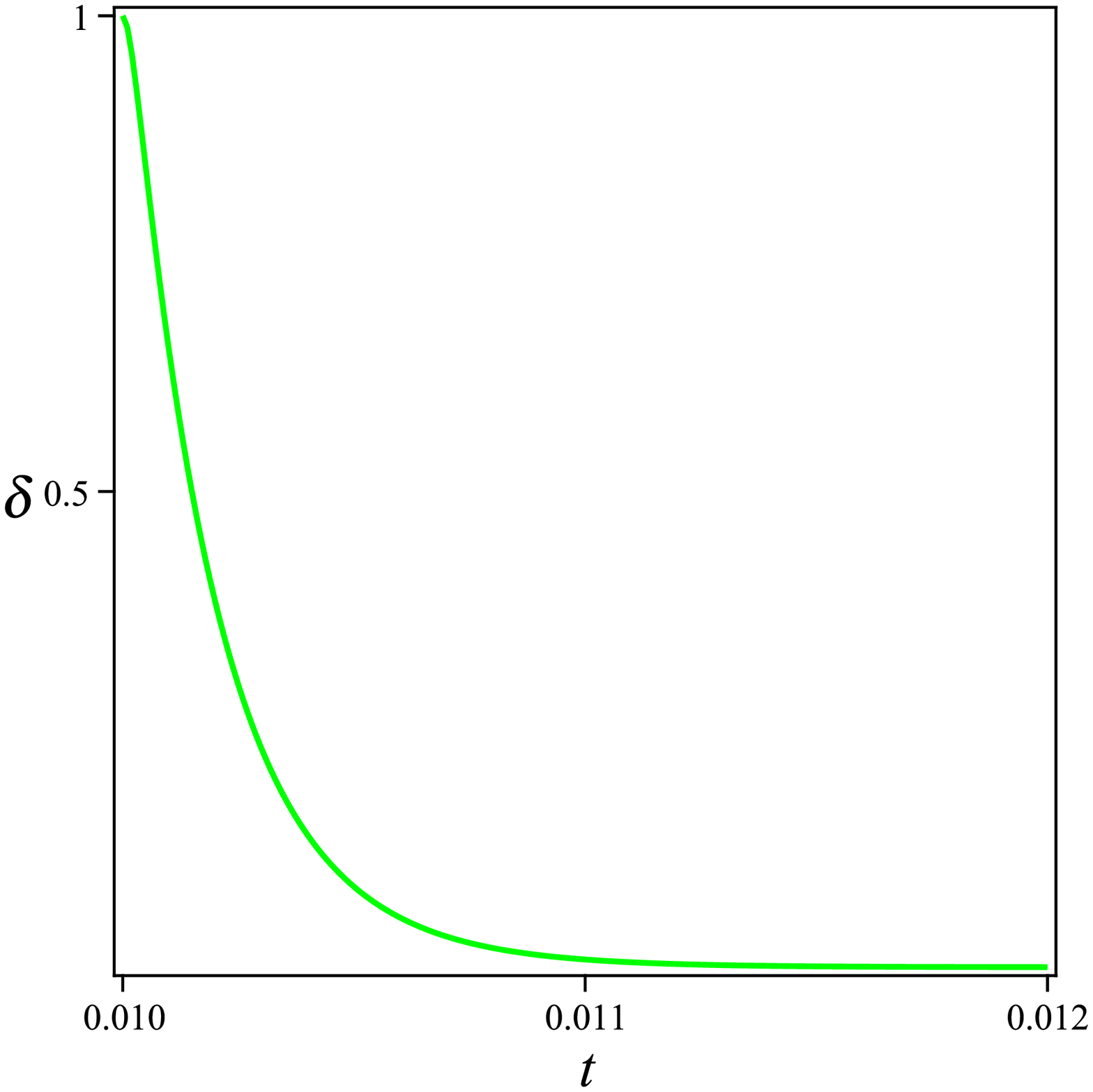}
 \end{array}$
 \end{center}
\caption{Time evolution of $\delta$ for the first exited case:
$\rho_{ecg}\gg\rho_{\phi}$ and $n=2$ at the early universe.}
 \label{fig:del2}
\end{figure}

\begin{figure}[h!]
 \begin{center}$
 \begin{array}{cccc}
\includegraphics[width=50 mm]{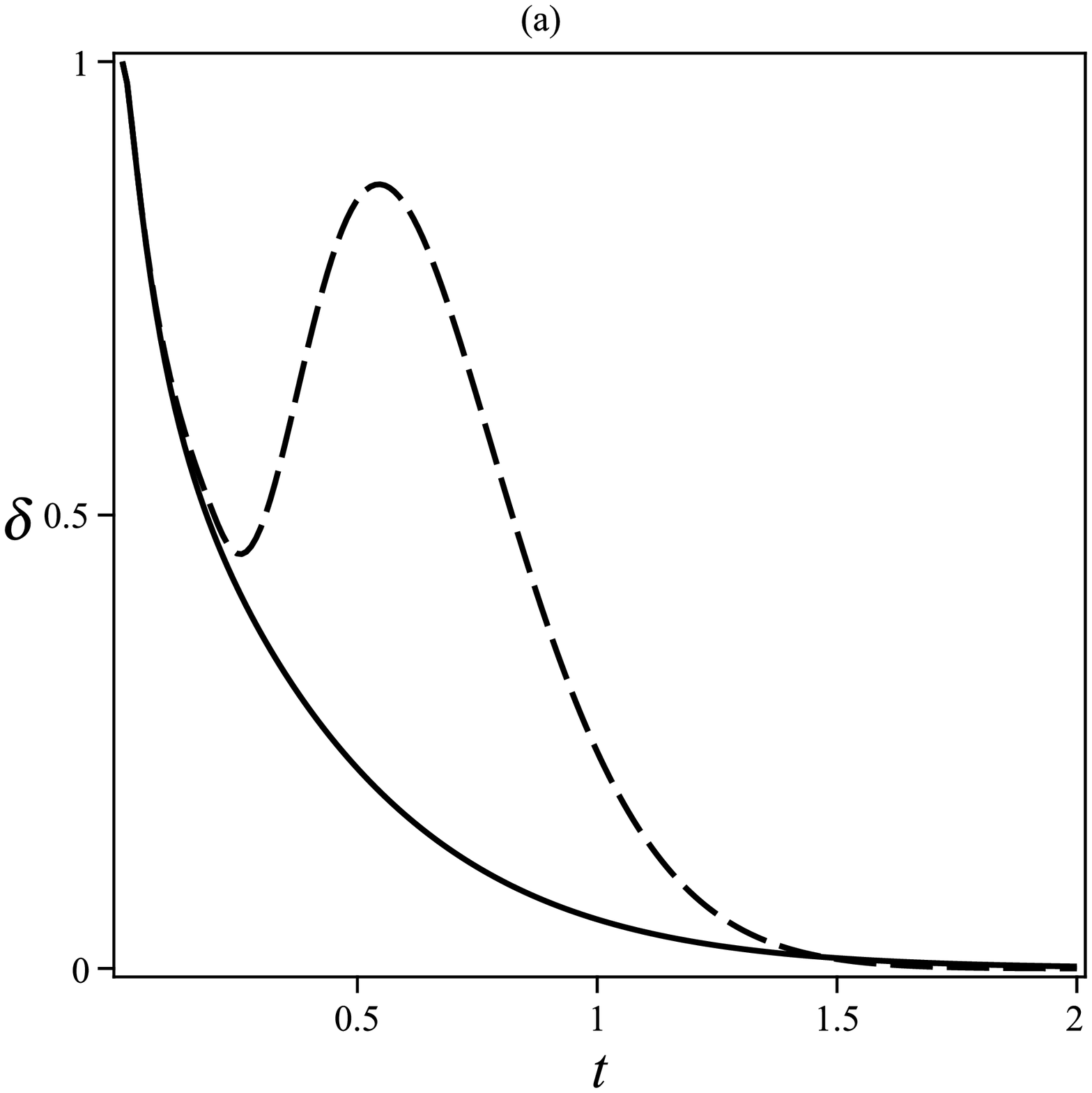}\includegraphics[width=50
mm]{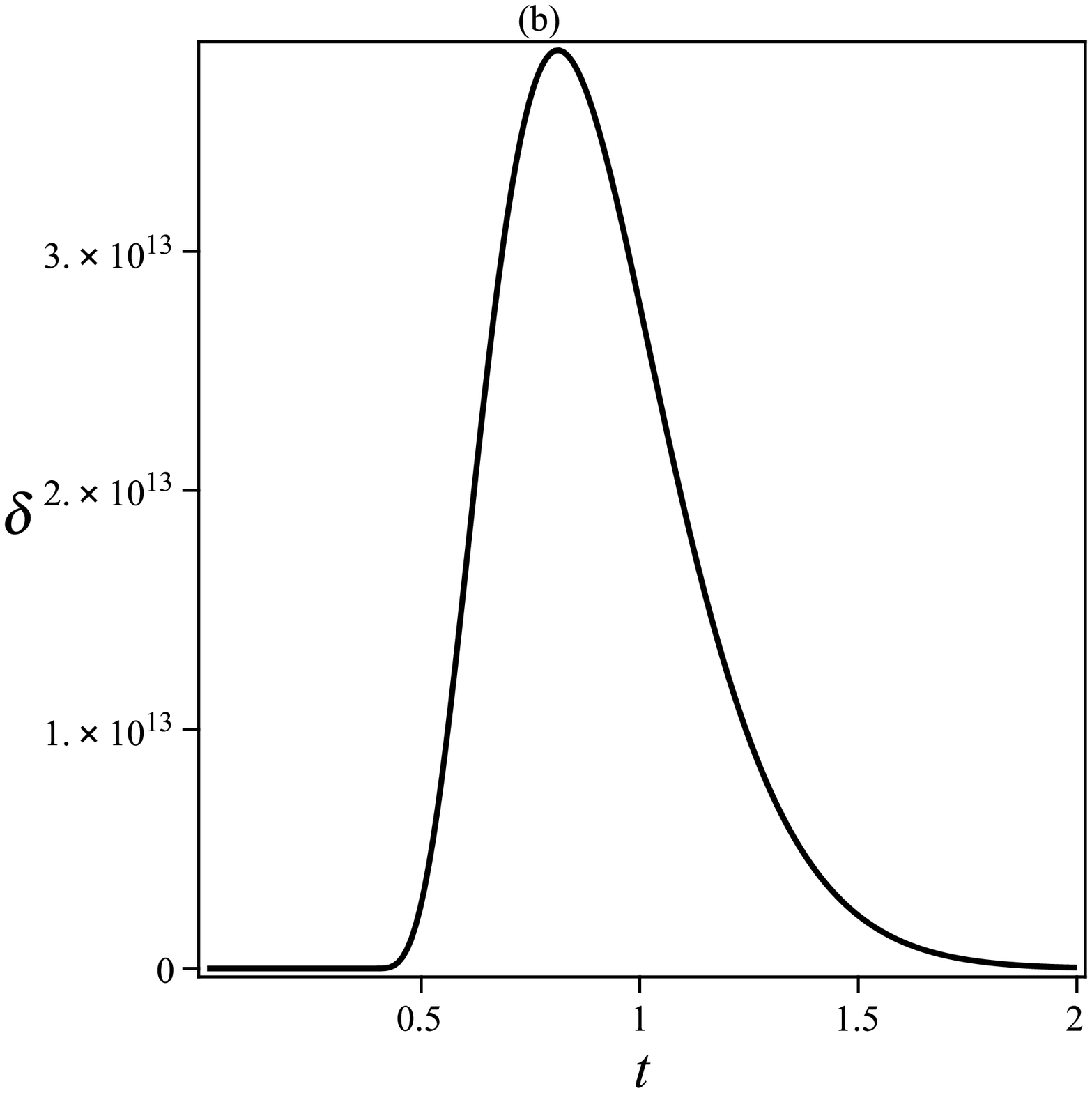}
 \end{array}$
 \end{center}
\caption{Time evolution of $\delta$ for the case of large $n$ and
$\rho_{ecg}\gg\rho_{\phi}$ at the early universe. (a) $b=0$ and $b=2$
represented by solid line and dashed line respectively. (b) the case of
$b=1$. }
 \label{fig:del3}
\end{figure}

Numerically, we can obtain behavior of $\delta$ for the cases which studied
in this paper. In the simplest case of $n=1$ we
see behavior of $\delta$ in the Fig. \ref{fig:del1}. It shows that perturbations grow
initially and vanished at the early universe. There is a maximum value at the
early universe.\\
Fig. \ref{fig:del2} shows behavior of $\delta$ corresponding to $n=2$ at the early
universe
under assumption of $\rho_{ecg}\gg\rho_{\phi}$. It is clear that the perturbations
vanished
suddenly at the
initial times. The maximum value is at the initial time exactly and
perturbation is totally decreasing function of cosmic time.\\
For the case of $n>2$ we can see from
the Fig. \ref{fig:del3} that value of interaction coefficient is important in evolution of
the perturbations. In the case of $b=1$, value of $\delta$ reaches to the large
number, however vanished at the early universe. In the case of without
interaction ($b=0$), perturbation $\delta$ is decreasing function of time and
vanished rapidly.\\
Therefore, we show that, density perturbations vanished at the early universe
and we can have stable model at the late time.\\
Our numerical results are base on
the simplest cases with $k=0$, however we can obtain similar results for the
cases of $k\neq0$.\\\\

\subsection{Tesor-to-scalar ratio}

\begin{figure}[h!]
 \begin{center}$
 \begin{array}{cccc}
\includegraphics[width=50 mm]{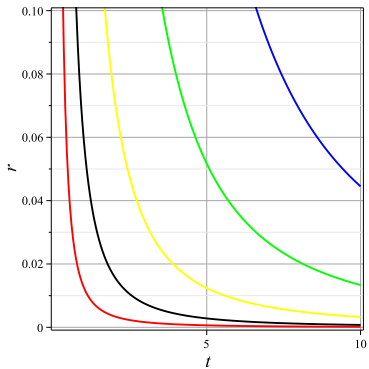}
 \end{array}$
 \end{center}
\caption{The tensor to scalar ratio in terms of cosmic time with $B=A_{n}=1$, $a_{1}=a_{2}=10$, $b=m=0.1$,
and $\alpha=0.5$. $n=1$ blue, $n=2$ green, $n=3$ yellow, $n=4$ black, $n=5$ red.}
 \label{fig:r}
\end{figure}

We can use our solutions to investigate the tensor
to scalar ratio and compare it with recent observational data \cite{P111}. We follow the
Ref. \cite{P112} to write,
\begin{equation}\label{tensor}
r\approx 32\left(\frac{\dot{H}}{H\dot{\phi}}\right)^2.
\end{equation}
We use energy density given by the equation (\ref{ansatz}) to find behavior of the tensor
to scalar ratio. Results of our numerical analysis represented in the plots of the Fig.
\ref{fig:r}. For simplicity we take $B=a_{1}=a_{2}=A_{n}=1$ and $\alpha=0.5$, hence the only free
parameter is $m$ (see Eq. (\ref{ansatz})). We can see that higher $n$ are closest to the observational
data.\\
We can find approximately similar behavior for bouncing solution given by ansatz
(\ref{s55}). In
this case also having agreement with KBP observation \cite{P111} and obtaining $r < 0.09$
needs to include higher $n$.

We can find approximately similar behavior for bouncing solution given by ansatz
(\ref{s55}). In
this case also having better agreement with $r < 0.09$ condition for higher $n$.

\section{Conclusion}
In this paper, we proposed a cosmological model based on phantom dark energy
which interacts with the extended Chaplygin gas. Therefore, we have two-component
fluid universe with possibility of interaction between components. First of all we introduced our model and wrote basic equations.
Then we discussed about solutions for the several cases and found inflationary and bouncing solutions. Numerically, we found behavior of the Hubble parameter at the late time for the general case and discussed briefly about the special case of $n=1$ which recovered modified Chaplygin gas. At the early universe we discussed cases of $n=2$, $n=2$ and $n>2$ separately and found agreement with the expected behavior of cosmological parameters. In the case of $n=2$ we reconstructed phantom potential and obtained behavior of phantom field as a function of cosmic time.\\
We also analyzed density perturbation of the model. By using squared sound speed we investigated stability of the model. In the case of $n = 1$ we found that our model is stable at the initial time. We have shown that the case of non-interacting yields to constant sound speed. By increasing
strength of interaction, the sound speed decreased and yields to a constant at the late time.  We found some instabilities at the early universe of the case $n=2$ depend on value of the interaction strength. However, instabilities vanished soon and our model will be stable. For the general case of $n>2$, we have seen that presence of interaction makes some instabilities at the initial times.  However, for the small
values of interaction coefficient we have stable model at the initial stage. Finally, comparing with recent observations of KBP \cite{P111} suggests that presence of higher order terms of the extended Chaplygin gas are necessary to obtain agreement with observations.\\\\

\begin{acknowledgments}
I would like to thank E. Saridakis and E.O. Kahya for useful comments and discussions.
\end{acknowledgments}

\end{document}